\documentclass[superscriptaddress,showpacs,amsmath,amssymb,prb,twocolumn]{revtex4}
\usepackage{graphicx}
\usepackage{dcolumn}
\usepackage{bm}
\usepackage{color}
\usepackage{amsmath}
\usepackage{amssymb}

\def\nb#1{}
\def\nba#1{}               

\begin{document}

\title{Quantitative nanostructure characterization using atomic pair distribution functions obtained from laboratory electron microscopes}

\author{Milinda Abeykoon}
\affiliation{Condensed Matter Physics and Materials Science Department, Brookhaven National Laboratory}
\author{Christos~D. Malliakas}
\affiliation{Department of Chemistry, Northwestern University}
\author{Pavol Juh\'{a}s}
\affiliation{Department of Applied Physics and Applied Mathematics, Columbia University}
\author{Emil~S. Bo\v{z}in}
\affiliation{Condensed Matter Physics and Materials Science Department, Brookhaven National Laboratory}
\author{Mercouri~G. Kanatzidis}
\affiliation{Department of Chemistry, Northwestern University}
\affiliation{Chemistry Department, Argonne National Laboratory}
\author{Simon J. L. Billinge}
\affiliation{Condensed Matter Physics and Materials Science Department, Brookhaven National Laboratory}
\affiliation{Department of Applied Physics and Applied Mathematics, Columbia University}

\begin{abstract}
\noindent Quantitatively reliable atomic pair distribution functions (PDFs) have been obtained from nanomaterials in a straightforward way from a standard laboratory transmission electron microscope (TEM).  The approach looks very promising for making electron-derived PDFs (ePDFs) a routine step in the characterization of nanomaterials because of the ubiquity of such TEMs in chemistry and materials laboratories.  No special attachments such as energy filters were required on the microscope.  The methodology for obtaining the ePDFs is described as well as some opportunities and limitations of the method.
\end{abstract}



\maketitle

\section{Introduction}
One of the great challenges of nanoscience is to obtain the quantitative structures of nanoparticles~\cite{billi;s07,jadzi;s07}.  The atomic pair distribution function (PDF) method has recently emerged as a powerful tool for doing this~\cite{egami;b;utbp03,juhas;n06,billi;jssc08,petko;cm06,page;chpl04,kodam;aca06,neder;pssc07,masad;prb07,petko;jpcc07}, but obtaining the required high quality diffraction data to high momentum transfer with good statistics generally requires synchrotron x-ray or spallation neutron data from a national user facility.  Here we show that data  of sufficient quality for quantitative analysis of nanoparticle structure using the PDF can be obtained from transmission electron microscopes (TEM) available at many research institutions.   Quantitative structural models were applied to PDFs of several nanoparticle systems showing that electron PDFs can be modeled with the powerful emerging modeling tools for studying PDFs in general~\cite{farro;jpcm07,neder;b;dsadss08,tucke;jpcm07,cerve;jac10}. This approach complements medium and high resolution imaging methods for studying nanoparticles in the TEM.  The ease of data collection and ubiquity of TEMs will make this an important tool in the characterization of nanostructured materials.

A challenge when using electrons as a probe is that they scatter strongly~\cite{cowle;mic04,cowle;b;edt92} and not according to the weak scattering kinematical scattering equations on which the PDF analysis is based~\cite{debye;ap15,warre;b;xd90}. This would appear to rule out electrons as a source of diffraction data for PDFs except in the cases of very dilute, such as gas-phase~\cite{schoo;nl05}, samples.
However, kinematical, or nearly kinematical, scattering is obtained from electrons when sample volumes are sufficiently small that multiple scattering events are not of high probability before the electrons exit the sample (typically a few nm of thickness), or when the scattering from the samples is highly incoherent, for example the scattering from amorphous materials and away from zone axes in a crystal~\cite{weiri;b;ecnafsdonm06}.  In these latter cases there is still significant multiple scattering, but it is sufficiently incoherent that it can be treated as a background and subtracted and the resulting coherent signal can be treated kinematically.  This has been discussed in detail in a number of publications~\cite{cocka;armr07,ansti;u88,ankel;zna05,karle;pnas77}. This is used in the rapidly growing field of electron crystallography~\cite{weiri;b;ecnafsdonm06}, and has been demonstrated in previous work of electron diffraction (ED) from glasses and amorphous materials~\cite{moss;prl69,cocka;aca88,cocka;armr07,hirot;mcp03,noren;jac99}, though little has been done in the way of quantitative modeling in those studies.  In these respects, the study of small nanoparticles is particularly favorable.  The samples are inherently thin, limited to the diameter of the nanoparticles when they are dispersed as a sub-mono-layer on a holey carbon support, and the structure is typically less coherent than from crystals because of the finite size effects that significantly broaden Bragg peaks and the often lower symmetries of nanoparticle structures due to surface and bulk relaxations.  In fact fortuitously, the scattering is most kinematical precisely for the small nanoparticles ($< 10$~nm) that are most beneficially studied using PDF methods~\cite{egami;b;utbp03,juhas;n06,billi;jssc08,petko;cm06,page;chpl04,kodam;aca06,neder;pssc07,masad;prb07,petko;jpcc07}.

Here we show how to obtain PDFs from a normal transmission electron microscope (TEM) found in many research labs.  We find that the resulting electron PDFs (ePDFs) can be modeled to extract quantitative structural information about the local structure using PDF refinement programs such as PDFgui~\cite{farro;jpcm07}.  This opens the door to broader application of PDF methods for nanostructure characterization since TEM is already a routine part of the nanoparticle characterization process~\cite{wang;jpcb00,won;jap06}.  With this development, as well as obtaining low and high resolution TEM \emph{images} of nanoparticles, quantitative structural information, similar to that normally obtained from a Rietveld refinement~\cite{rietv;jac69,young;b;trm93} in bulk materials, is also available from nanoparticles with little additional effort.  This approach also complements high resolution TEM by getting an average signal from a large number of nanoparticles rather than giving information from a small part of the sample that may not be representative.  The fact that the real-space images and the diffraction data suitable for structural analysis can be obtained at the same time and from the same region of material is also a large advantage, resulting in more complete information for the characterization of the sample.  In some cases, the small quantity of material required for ePDF, compared to x-ray and neutron PDF measurements (xPDFs and nPDFs, respectively), may also be a major advantage, as well as the ability to study thin films.

\subsection*{Theoretical Background}

\begin{figure}[tb]
\center
\includegraphics[scale=1, angle=0]{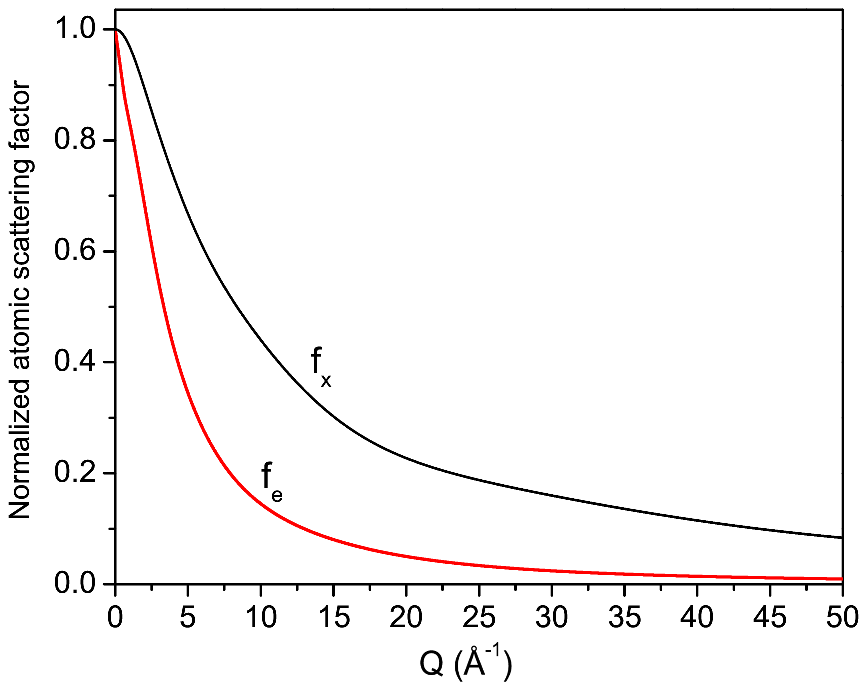}
\caption { A comparison between normalized (to f(0)) x-ray and electron form factors, $f_{x}$(Q) and $f_{e}$(Q) of
Au.} \label{fig:Au_formfactors}
\end{figure}

\noindent The Fourier transform of X-ray or neutron powder diffraction data yields the PDF, $G(r)$ according to~\cite{farro;aca09}
%
\begin{eqnarray}
\label{eq:PDF}
G(r)=(2/\pi)\int_{Q_{\min}}^{Q_{\max}}Q[S(Q)-1]\sin(Qr)dQ,
\end{eqnarray}
%
where the structure function, $S(Q)$, is the properly normalized powder diffraction intensity and $Q$, for elastic scattering, is the magnitude of the scattering vector, $Q = 4\pi \sin(\theta)/\lambda$.\cite{warre;b;xd90,egami;b;utbp03} The PDF is also related to the atomic structure through
%
\begin{eqnarray}
\label{eq:PDF2} G(r)=\frac{1}{N r}\sum_{ij} \frac{f_{i}(0)f_{j}(0)} {\langle f(0) \rangle^{2}}
\delta(r-r_{ij})-4\pi r \rho_{o},
\end{eqnarray}
%
where the sum goes over all pairs of atoms \emph{i} and \emph{j} separated by $r_{ij}$ in the model. The form factor of atom \emph{i} is $f_{i}(Q)$ and $\langle f(Q) \rangle$ is the average over all atoms in the sample. In equation~\ref{eq:PDF2}, the scattering factors are evaluated at $Q=0$, which in the case of x-rays is the atomic number of the atom. The double sums are taken over all atoms in the sample.
For a multicomponent system, $S(Q)$ can be written in terms of the concentrations, $c_i$, of the atoms\cite{warre;b;xd90,egami;b;utbp03}
%
\begin{equation}
\label{eq:SQ} S(Q)=1+\frac{I(Q)-\sum c_{i}|f_{i}(Q)|^{2}}{\big|\sum
c_{i}f_{i}(Q)\big|^{2}}.
\end{equation}
%
In the case of electrons as a probe, the equations are the same, providing the scattering can be treated kinematically~\cite{cowle;b;edt92}; however, the form-factor must be that appropriate for electrons, $f_{e}(Q)$, which is the Fourier transform of the electronic potential distribution of an atom.  Note that in the electron diffraction literature, it is common to use $s = 2\sin(\theta)/\lambda=Q/2\pi$ instead of $Q$ for the independent variable in the scattering. The electron form factor, $f_{e}(Q)$, is different to, but closely related to, the x-ray form factor of the same atom, $f_{x}(Q)$, which is the Fourier transform of the electron density.  A useful relationship between $f_{e}(Q)$ and $f_{x}(Q)$ is~\cite{cowle;b;edt92}
%
\begin{equation}
\label{eq:fe}
f_{e}(Q)=\frac{m_e e^{2}}{2\hbar^{2}}\left(\frac{Z-f_{x}(Q)}{Q^2}\right),
\end{equation}
%
where $m_e$ and $e$ are the mass and charge of the electron, respectively, $\hbar$ is Plank's constant, and $Z$ the atomic number. This equation does not give a definite value for $f_{e}(Q)$ at $Q =0$, but $f_{e}$(0) can be calculated by extrapolation or by using
%
\begin{equation}
\label{eq:f0}
f_{e}(0)=4\pi^{2}\frac{m e^{2}}{3\hbar^{2}}(Z\langle r_e^{2}\rangle),
\end{equation}
%
where $\langle r_e^{2}\rangle$ is the mean square radius of the
electronic shell of the atom.\cite{cowle;b;edt92} Figure~\ref{fig:Au_formfactors} shows a comparison between x-ray and electron form factors, $f_{x}$(Q) and $f_{e}$(Q) of Au.


In the case of single crystal ED, a rule of thumb is that when the crystal thickness is greater than $\sim300$-400~{\AA}, data reduction must be done based on the dynamical diffraction theory which assumes the presence of coherent multiple scattering components of electrons~\cite{cowle;b;edt92}. Depending on the energy of the electrons, this thickness limit may even fall below these numbers in the presence of heavy elements~\cite{cowle;b;edt92},   and in the case of electron powder diffraction, the average thickness of crystallites in the specimen should also be less than a few hundred {\AA}ngstr\"oms to avoid dynamical scattering effects.~\cite{cowle;b;dp95} Coherent multiple scattering changes the relative intensities of Bragg peaks from the kinematical structure factor values, redistributes intensity to the weaker peaks at higher values of $Q$~\cite{cocka;aca88} and can allow symmetry disallowed peaks to appear in the pattern.   If $\Gamma$ is the elastic mean free path of the electron, it has been shown that a PDF determined from a polycrystalline Pt sample does not affect the positions of the PDF peaks for D/$\Gamma\le 5$, where D is the particle size, but it does affect the determination of coordination numbers~\cite{ansti;u88}.  Here we show that model fits may be good, even in the presence of significant multiple scattering, while refined thermal factors are underestimated, though it is desirable to optimize experimental conditions such as to minimize multiple scattering.

Incoherent multiple scattering can be observed in ED patterns in the form of increased background~\cite{cowle;b;edt92} which does not affect the relative intensities of the Bragg peaks.  This is why, in the case of a less coherent structure, dynamical scattering effects are less important.

In this study, all the specimens used were nanosized samples: thin films, discrete nanoparticles, or agglomerates of nanoparticles. In this case, the hope was that multiple scattering would not introduce undue aberrations into the kinematical diffraction pattern and a reliable PDF will result.  We found this to be largely true with an exception we discuss below.

\section{Experimental}
Nanocrystalline thin film, or dispersed nanoparticulate samples, were distributed on a holey carbon grid and ED data taken with a short camera length, to give the widest $Q$-range, and a relatively large beam-size (2-5~$\mu$m diameter) on the sample, to obtain the best possible powder average.  To improve the powder average different regions of the sample were illuminated by translating the sample under the beam.  In other respects, the TEM was used in a standard configuration using a CCD detector and no energy filtering and operated at 200~keV (wavelength, $\lambda = 0.025079$~{\AA}).

All selected area ED experiments were carried out at room temperature on a Hitachi H8100 200~KeV transmission electron microscope equipped with a Gatan Orius SC600 CCD Camera (24mm x 24 mm active area). Typical exposure time per frame was around 0.3s. Formvar coated 300 mesh copper grids (Electron Microscopy Sciences) stabilized with an evaporated carbon film were used to support the metallic films and nanoparticles. Deposition of gold on the carbon coated side of the TEM grid was performed with a Denton Vacuum DeskIII sputterer and gave a uniform film. The thickness of the film was measured in real time during the sputtering process with the aid of a thickness monitor (Maxtek, Inc TM-350). Part of the grid was masked during the deposition and this masked area was used to extract the diffraction intensity of the support. No differences were found in the diffraction intensity data of the background (carbon and polymer films) between different grids. Deposition on the side of the grid that was coated with the polymer gave Au nanoparticles with a wide range of sizes up to $\sim 100$~nm. NaCl nanoparticles were deposited on a TEM grid by a radio-frequency thermal evaporation method.

\nb{can you write a sentence about where the Au particles came from? -- I will ask Christos }
For comparison, x-ray measurements on Au nanoparticles were carried
out in the rapid acquisition mode (RAPDF)~\cite{chupa;jac03} using a
Perkin Elmer amorphous silicon 2D detector at X7B beamline of National
Synchrotron Light Source (NSLS) at Brookhaven National Laboratory
(BNL). Nanoparticles in ethanol solution  were loaded in a 1~mm diameter kapton tube sealed at both ends, and mounted perpendicular to the x-ray beam.  The data were collected at room temperature using the x-ray energy of $\sim$38~keV ($\lambda = 0.3196$~{\AA}). The data were collected in a multiple 4~s exposures for a total collection time of 5~min.

To calibrate the conversion from detector coordinates to scattering angle, it is necessary to measure the ED pattern from a standard of known lattice parameters.  The software for reducing the data to 1D, Fit2D~\cite{hamme;hpr96} uses this to optimize the effective sample-detector distance, find the center of the Scherrer rings on the detector, and correct for aberrations such as any deviation from orthogonality of the detector and the scattered beam.  Typical standards used by the program are Al$_{2}$O$_{3}$, CeO$_{2}$, LaB$_{6}$, NaCl  and Si.  However, for the ED experiment it is necessary to have a nano-sample standard to obtain a good powder average. For this gold nanoparticles of diameter $\sim$100~nm were used and a literature value of 4.0782~\AA\ for the lattice parameter. The effective sample-detector distance depends on the settings of the magnetic lenses used in the microscope. We assumed that the energy of the electrons, 200~keV, is well known (resulting in $\lambda =0.025079$~{\AA}), though for the most accurate results the electron wavelength should be calibrated using standard methods.
Once these calibration quantities are known, they are fixed and the same values are used to convert the sample data.  From this perspective it is essential that the sample is measured under identical conditions as the standard, including camera length and focus. We found that even scanning around a sample to find a different viewing area resulted in a small variation in the position on the detector of the center of the resulting diffraction pattern.  It was thus necessary to run a separate calibration run on each diffraction pattern to determine the center of the rings, while keeping the camera-length from the Au calibration.

\section{Data Analysis}

The 2D ED images were read and integrated into 1D powder diffraction
patterns, after masking the missing beam stop
region. The data have to be further processed to obtain the
PDF. Corrections were applied to the raw data to account for
experimental effects and properly normalized and divided by ($\langle
f_e(Q) \rangle^{2}$)~\cite{egami;b;utbp03}, resulting in the total
scattering structure function, $S(Q)$.  The kernel of the Fourier
transform is the reduced structure function, $F(Q)=Q[S(Q)-1]$.  We
used a home-written program, PDFgetE, to carry out these steps. The
PDF is then straightforwardly obtained as the Fourier transform of
$F(Q)$ according to Eq.~\ref{eq:PDF}, which is also carried out in
PDFgetE. Once the PDFs are obtained, they can be modeled using existing PDF
modeling programs. Here we used  PDFgui~\cite{farro;jpcm07}.

\section{Results}

\begin{figure}[tb]
\center
\includegraphics[scale=.65,angle=0]{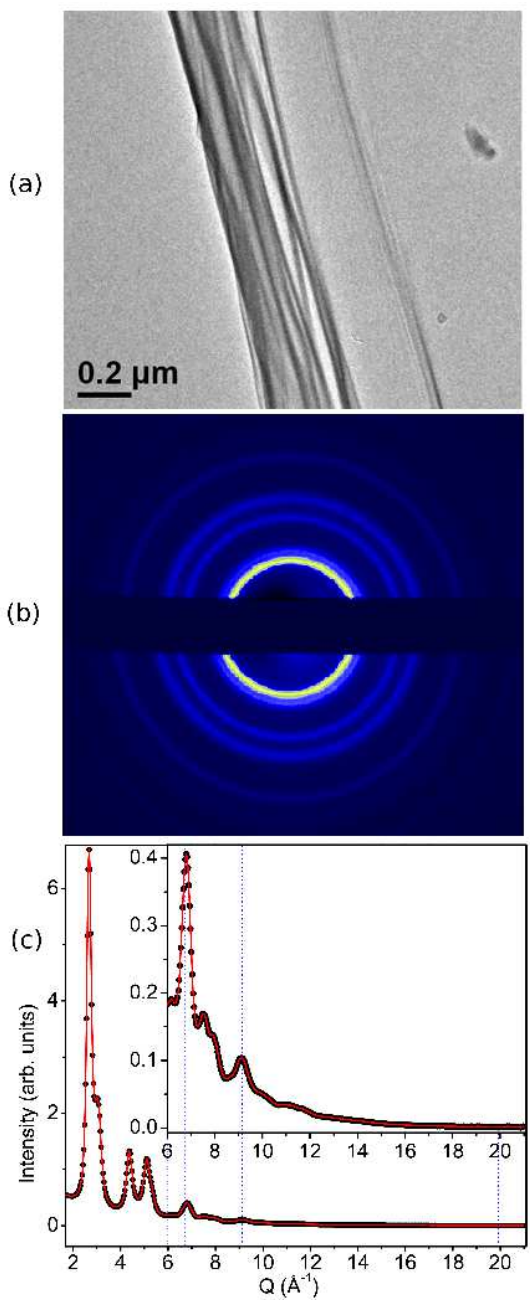}
\caption {(a) A TEM image of the 2.7 nm thick Au film used for ED. (b)~A false-color 2D ED pattern collected on this sample using 200~keV electrons.  Lighter colors indicate higher intensity.  The black bar across the middle of the image is the shadow of the beam-stop. (c) 1D Au electron powder diffraction pattern obtained by integrating around the rings in ~\ref{fig:Au_image_ED} (b). The inset shows the high~$Q$ region of the ED pattern on an expanded
$y$-scale. The dotted lines are guides to the eye. } \label{fig:Au_image_ED}
\end{figure}

A low resolution TEM image of the 2.7~nm thick Au film is shown in Fig.~\ref{fig:Au_image_ED}(a).  The film is uniform and featureless in the image, but a region at the edge of the film was selected for imaging so that the edge of the film gives a visual cue to its presence. An ED pattern from a position away from the edge of the film is shown in Fig.~\ref{fig:Au_image_ED}(b). We can see a series of concentric circles due to the Scherrer powder diffraction rings in transmission geometry. The resulting 1D ED pattern, obtained by integrating around the rings in the 2D pattern is shown in Fig.~\ref{fig:Au_image_ED}(c). Broad diffuse features are observed consistent with the nanocrystallinity of the sample.  Weak features are clearly evident up to $Q=12$~{\AA}$^{-1}$  (Fig.~\ref{fig:Au_image_ED}(c) inset), but less apparent beyond that point.

\begin{figure}[tb]
\center
\includegraphics[scale=1.1,angle=0]{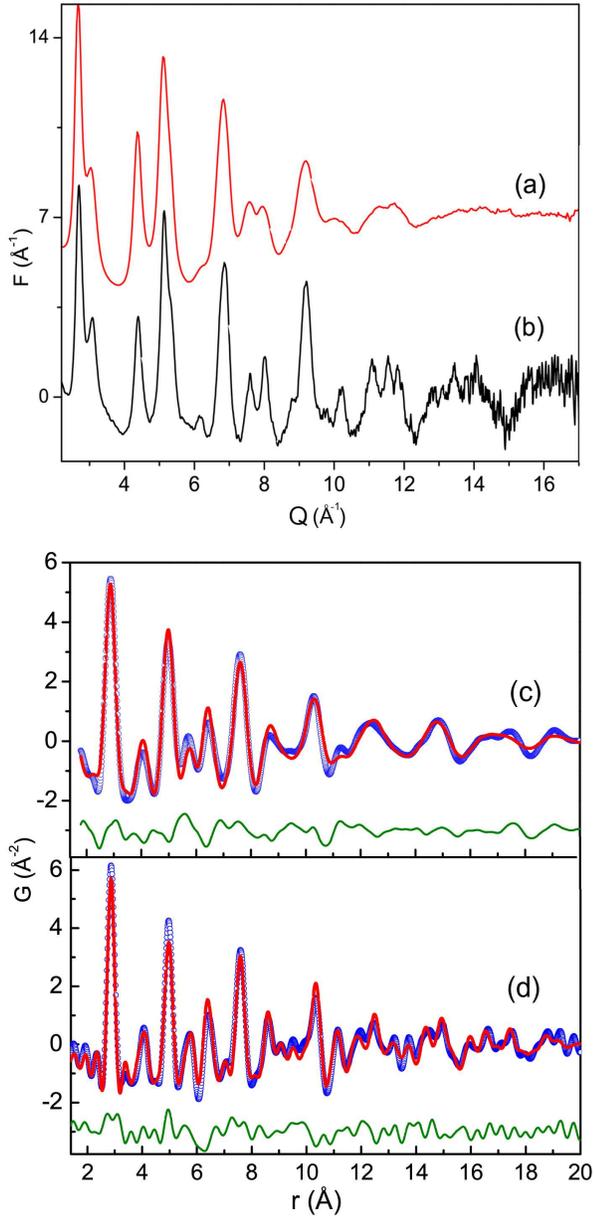}
\caption {(a) Reduced structure function, $F(Q)$, of Au obtained from the integrated ED pattern in Fig.~\ref{fig:Au_image_ED}(c),
(b) An $F(Q)$ of Au nanoparticles calculated from an XRD pattern collected at X7B at the NSLS.  (c) Au bulk structure model fit to
  the resulting ePDF from ~\ref{fig:Au_xFq_eFq_2}(a). (d)  Au bulk structure model fit to the resulting xPDF
  from ~\ref{fig:Au_xFq_eFq_2}(b). Observed and calculated PDFs are presented with blue circles and a solid red
  line respectively. The difference between observed and calculated is offset below (green solid
  lines). In both cases used Q$_{\max}$=15.25~{\AA}.}\label{fig:Au_xFq_eFq_2}
\end{figure}

The $F(Q)$ from the same data after correction is shown in Fig.~\ref{fig:Au_xFq_eFq_2}(a) and the resulting ePDFs   in Fig.~\ref{fig:Au_xFq_eFq_2}(c), with the calculated PDF from a model of the gold fcc structure plotted on top in red. For comparison, in Fig.~\ref{fig:Au_xFq_eFq_2}(b) and (d) we show the x-ray derived $F(Q)$ and xPDF, respectively.  Unfortunately this is not a direct comparison between identical samples. We were not able to collect x-ray data from the same film as the ePDF as it was too thin to get a sufficient signal in the x-ray measurement.

The structure functions of the electron and x-ray data (Fig.~\ref{fig:Au_xFq_eFq_2}(a) and (b), respectively) are clearly highly similar.  Features in the $eF(Q)$ are broader than the x-ray case but the features are all recognizable and have the correct relative intensities.  Likewise, the e- and xPDFs (Fig.~\ref{fig:Au_xFq_eFq_2}(c) and (d), respectively) are highly similar, with the features in the ePDF of the nanoscale film being broader.  The quality of the fits is comparable for both the ePDF and xPDF curves, with the ePDF giving a slightly lower (better) agreement factor. The refined parameters are presented in Table~\ref{tab:Au_Results}.  The breadth of the ePDF peaks are accommodated in the model by giving gold very large atomic displacement parameters (ADPs), twice as large as those in the x-ray measured gold nanoparticles that are already large.  This indicates the presence of significant atomic scale disorder in the film and is not coming from the ePDF measurement itself.  This is discussed in greater detail below.

These results clearly demonstrate that quantitatively reliable ePDFs can be obtained from nanocrystalline materials in a standard laboratory TEM.  The counting statistics from the electron data compare favorably to those from the x-ray measurements (Fig.~\ref{fig:Au_xFq_eFq_2}(a) and (b)), despite the much shorter measurement time, suggesting that ePDF determination could become a useful general characterization tool during nanoparticle synthesis. Two effects are clearly evident in the $Q$-space data: low
$Q$-space resolution and the rapid diminishing of the amplitude of scattered features with
increased $Q$.  The latter is likely to reflect real differences in the samples, with the range of structural coherence being lower in the gold film than the gold nanoparticles used in the x-ray experiment.  The lower $Q$-space resolution could be either a sample or a measurement effect but this cannot be disentangled without having a well characterized, kinematically scattering, nanoparticle standard for ED, which doesn't currently exist. The sputtered gold film has an fcc gold structure, like the bulk, but with significantly more disorder and a nanometer range for the structural coherence.

The ED data were taken with a standard CCD camera and no filtering of inelastically scattered electrons. This is the most straightforward protocol for data collection as it is the standard setup in most laboratory TEMs.  It is expected to result in lower quality PDFs than those measured with energy filtered electrons because of the higher backgrounds due to inelastically scattered electrons~\cite{cocka;aca88}.  ED data collected with an image plate detector  are also expected to be higher quality due to the low intrinsic detector noise and better dynamic range of that detector technology.  Thus, the resulting PDF shown in Fig.~\ref{fig:Au_xFq_eFq_2}(c) represents the baseline of what is possible without specialized instrumentation. The resulting $F(Q)$ shows excellent signal to noise up to the maximum accessible $Q$-range of 17~\AA$^{-1}$,  as evident in Fig.~\ref{fig:Au_xFq_eFq_2}(a).

\begin{figure}[tb]
\center
\includegraphics[scale=.65,angle=0]{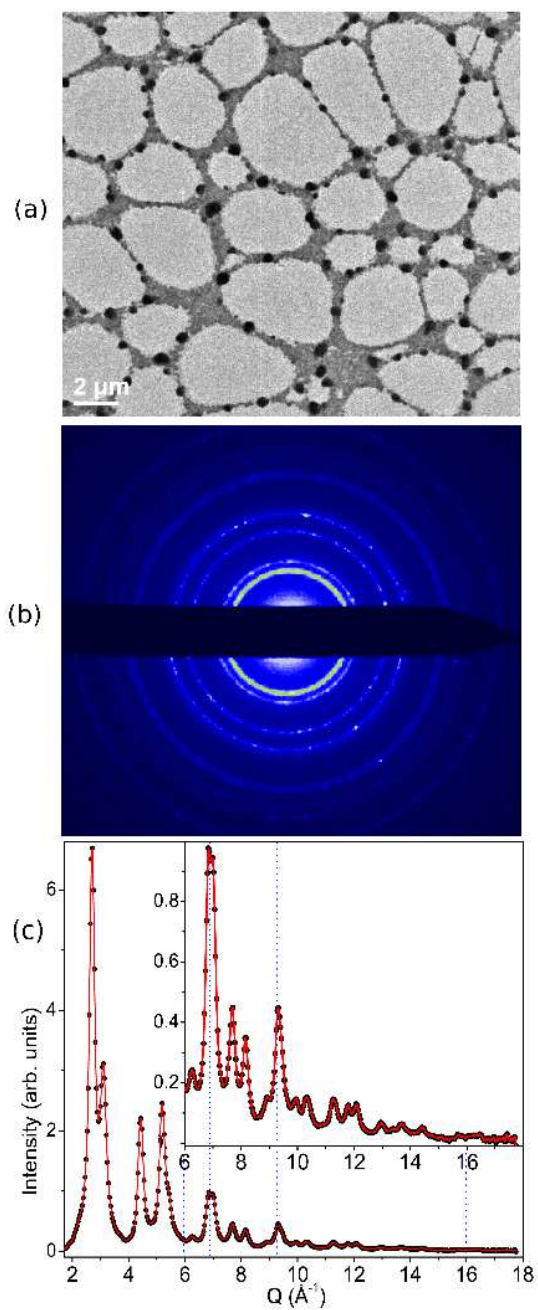}
\caption {(a) A TEM image of  $\sim$ 100~nm Au nanoparticles used for ED. Black dots are the nanoparticles on the grid and the large white areas are the holes in the grid.  (b)  A background
subtracted ED image, collected from the same region of the sample using 200~keV electrons. (c) The 1D  ED pattern
obtained by integrating around the rings in ~\ref{fig:Au_thick_image_ED}(b). The inset shows a magnified region of
the integrated ED pattern as indicated by the dotted lines. This ED pattern clearly suffers from multiple scattering
due to the thickness of the sample.} \label{fig:Au_thick_image_ED}.
\end{figure}

\begin{figure}[tb]
\center
\includegraphics[scale=1.1,angle=0]{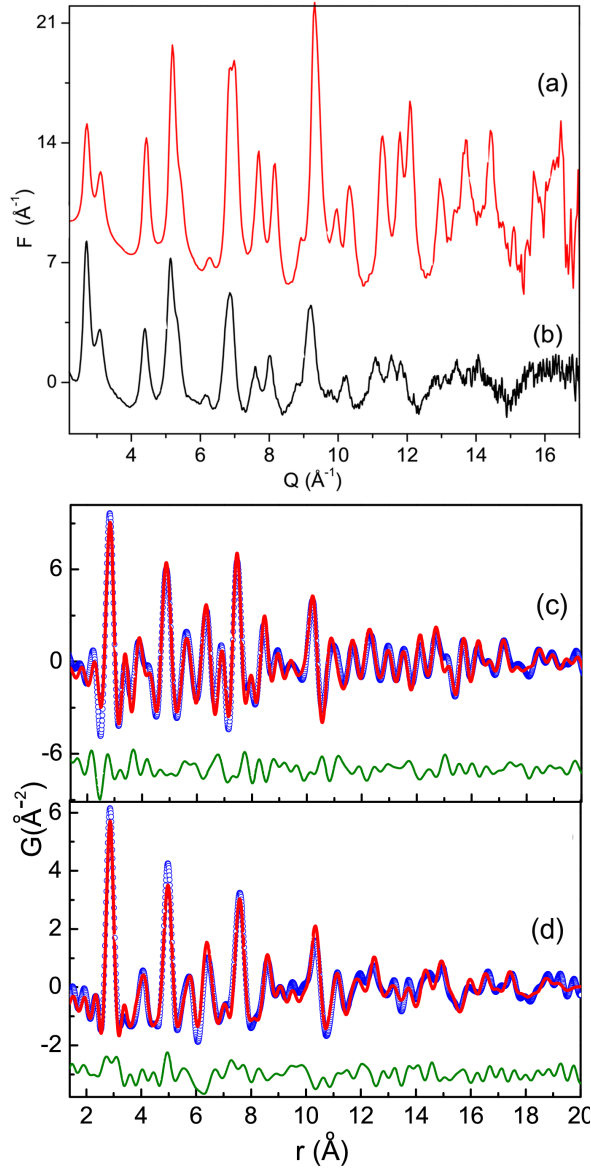}
\caption {(a) The reduced structure function, $F(Q)$, of Au calculated from the  integrated ED pattern in
  Fig.~\ref{fig:Au_thick_image_ED}(c), (b) An F(Q) generated from an integrated Au NP XRD pattern from a 2D data set collected at X7B at the NSLS. (c) Au bulk structure model fit
  to the resulting ePDF  from ~\ref{fig:f_comparison_2} (a). (d)  Au bulk structure model fit to the resulting xPDF
  from ~\ref{fig:f_comparison_2} (b). Observed and calculated PDFs are presented with blue circles and a solid red
  lines respectively. The difference between observed and calculated is offset below (green solid lines).}
\label{fig:f_comparison_2}
\end{figure}

To explore the size limits for Au NPs to scatter kinematically, we collected data from larger,  100~nm  Au nanoparticles, and the results are also given in Table~\ref{tab:Au_Results} and Figures~\ref{fig:Au_thick_image_ED} and~\ref{fig:f_comparison_2}.
Comparing the integrated 1D diffraction patterns of the large NPs and the thin Au film,
Figs.~\ref{fig:Au_thick_image_ED}(c) and~\ref{fig:Au_image_ED}(c), respectively, we see similar features, but in the case of the NPs, the amplitudes of the scattered intensities extend to much higher $Q$ values, as if there is a much smaller Debye-Waller factor for the data.   This can be clearly observed by comparing the e$F(Q)$ of the large nanoparticles in Fig.~\ref{fig:f_comparison_2}(a) with those from x-ray diffraction data, x$F(Q)$ in Fig.~\ref{fig:f_comparison_2}(b). The enhancement in the high-$Q$ features is large and is almost certainly due to significant coherent multiple scattering in this sample.  The resulting ePDF from the NPs has peaks that are correspondingly sharp compared to the thin film gold and the xPDFs of gold NPs.  Regardless of the presence of significant multiple scattering, a model was refined against the ePDF of the 100~nm Au nanoparticles to see the extent that the refined structural model parameters are affected. The structure refinement gave fits that were slightly worse but comparable in quality to the xPDF fits (see Table~\ref{tab:Au_Results}), $R_w=0.24$.  The refined values were similar also, except for much smaller atomic displacement parameters (ADPs), due to the artificially sharpened PDF peaks.  It is somewhat remarkable that, in this case, the dynamical scattering produces features in the $F(Q)$ with approximately the correct relative amplitude, but extending to much higher-$Q$.  Not only are the PDF peaks in the right position~\cite{ansti;u88}, but have the right relative amplitudes.  Gold may be a special case because the structure factors are all either ones or zero's.

This clearly shows that for a strong scatterer such as Au, 100~nm nanoparticles already give significant dynamical effects.  The resulting PDFs give useful semi-quantitative and qualitative information but the refined thermal parameters are not reliable.  Indeed, the effect of the multiple scattering to increase the real-space resolution by boosting the intensities of the high-$Q$ peaks makes the PDF peaks sharper with the result that bond-lengths can be extracted with greater precision from the ePDF data in this case.  When accurate PDF peak positions rather than quantitative peak intensities are desired, this could be a significant advantage of the ePDF method, for example, when looking for small peak splittings, or resolving peak overlaps, to aid in structure solution.

\begin{figure}[h!]
\center
\includegraphics[scale=.65, angle=0]{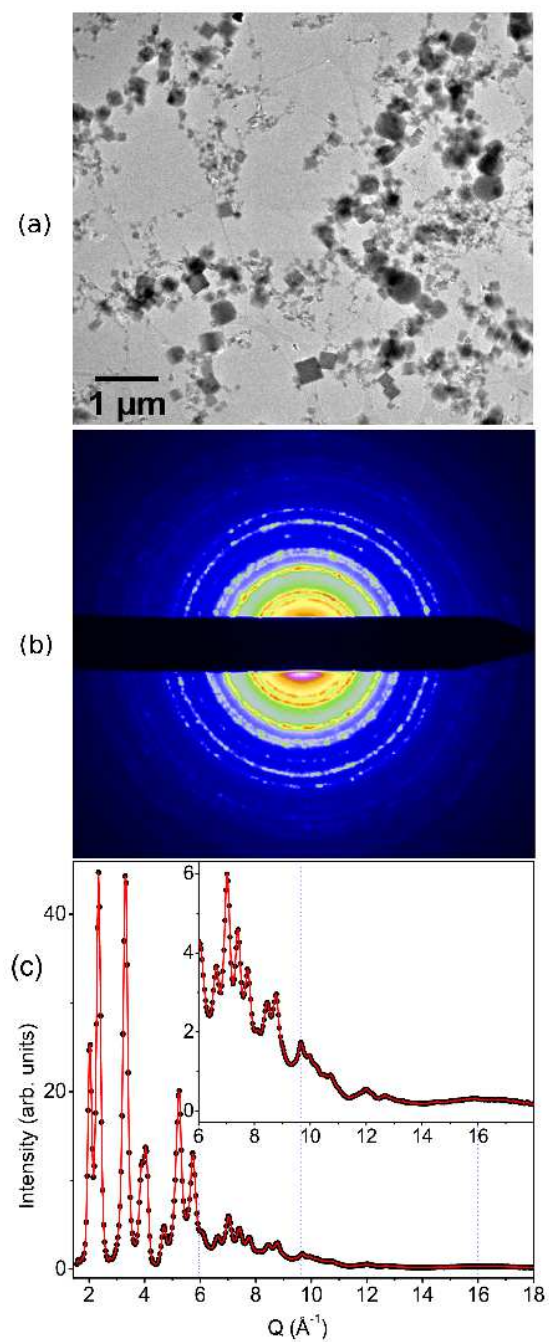}
\caption {(a) A TEM image of the NaCl film used for ED.  (b) A false-color 2D ED image collected on this sample using 200~keV electrons.  Lighter colors indicate higher intensity. The black bar across the middle of the image is the shadow of the beam-stop. (c) 1D ED pattern obtained by integrating around the rings in ~\ref{fig:NaCl_image_ED}~(b).  The inset shows the high~$Q$ region of the ED pattern on an expanded $y$-scale. The dotted lines are guides to the eye.} \label{fig:NaCl_image_ED}
\end{figure}

\begin{figure}[h!]
\center
\includegraphics[scale=1.1,angle=0]{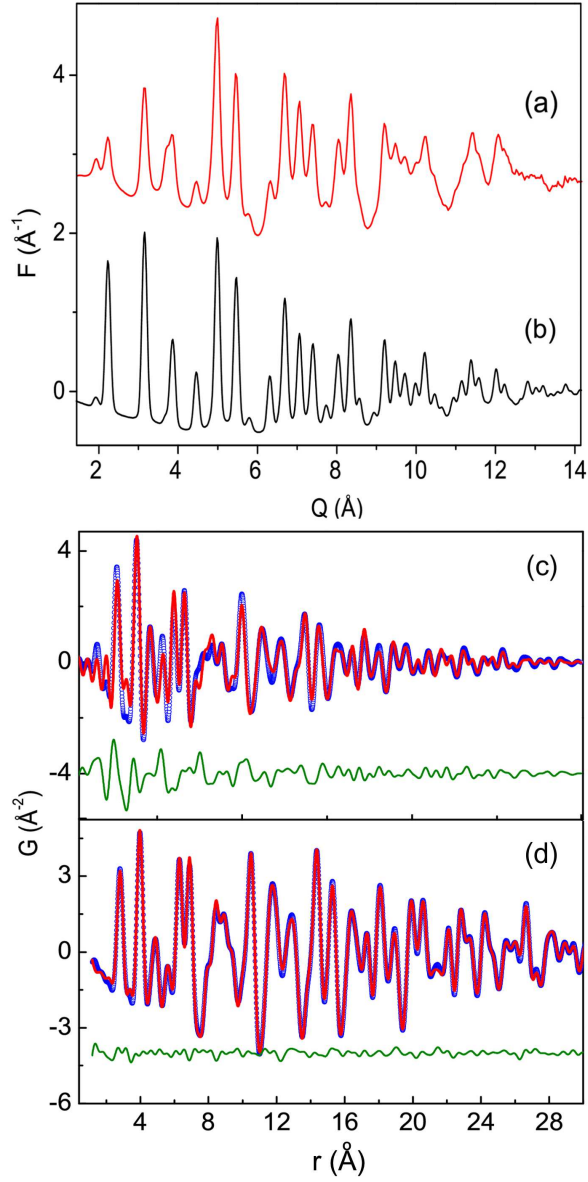}
\caption {(a) Reduced structure function, F(Q), of NaCl obtained from the  integrated ED pattern in
  Fig.~\ref{fig:NaCl_image_ED}(c).  (b) An F(Q) of NaCl calculated from an x-ray data set.  (c) NaCl bulk structure model fit to the resulting ePDF from ~\ref{fig:NaCl_FQ_Gr} (a). (d)  NaCl
  bulk structure model fit to the resulting xPDF from ~\ref{fig:NaCl_FQ_Gr} (b). Observed and calculated PDFs are presented with blue circles and solid red lines respectively. The difference between observed and calculated is offset below (green solid lines).  In both cases used Q$_{\max}$=13.6~{\AA}.} \label{fig:NaCl_FQ_Gr}
\end{figure}

A less trivial structure factor is obtained from binary compounds, such as the NaCl studied here.  The TEM image of the sample in Fig.~\ref{fig:NaCl_image_ED}(a) shows that it consists of nanoscale crystallites, some of which have a cubic habit and others that have no particular morphology.  The corresponding ED pattern in Fig.~\ref{fig:NaCl_image_ED}(b) shows clear and fairly uniform rings, with some spottiness from an imperfect powder average. Fig.~\ref{fig:NaCl_image_ED}(c) shows the integrated ED pattern. The $F(Q)$ and the resulting ePDF obtained from this data set is shown in Figs.~\ref{fig:NaCl_FQ_Gr}(a) and (c), respectively.  For comparison, an x$F(Q)$ and an xPDF obtained from a bulk crystalline NaCl sample is also shown in Fig.~\ref{fig:NaCl_FQ_Gr}(b) and (d).

The rock-salt structure model fits to the PDFs are shown in Fig.~\ref{fig:NaCl_FQ_Gr}(c) and (d) and the results are presented in Table~\ref{tab:NaCl_results}. The e- and xPDFs are qualitatively highly similar, with all features in the xPDF easily recognizable in the ePDF. Notably, the relative intensities of adjacent peaks are similar between the e- and xPDFs. Peaks in the ePDF die out in amplitude with increasing~$r$ more quickly, due to the broader features in the ED pattern. The overall quality of the fit to the ePDF is worse than the xPDF of bulk NaCl.  Refined lattice constants agree well within the experimental uncertainty. The ePDF refined thermal parameters are much smaller than those obtained from the x-ray data.  This is unlikely to be a real effect as both the x-ray and electron data were measured at room temperature, and it is rather implausible that the nanoparticulate samples have \emph{less} static structural disorder than bulk NaCl.  We therefore assume that this is the effect of  multiple scattering in the data, similar to that observed for large Au NPs.  Clearly, ADPs refined from ePDFs present a lower bound on actual sample ADPs.  They are accurate in the case where multiple scattering is negligible, but underestimate the thermal motions and static disorder in the presence of multiple scattering.

\section{Discussion and Conclusions}

The Au and NaCl examples establish that quantitatively, or semi-quantitatively, reliable PDFs can be obtained from nanomaterials using electron diffraction data obtained on a standard laboratory TEM, without the use of filtering.  Because of the ease and speed of collecting such data and the ubiquity of such instruments in chemistry and materials laboratories, if the barriers to data processing could be overcome making the whole process straightforward, this could become a broadly applicable standard and useful characterization method for nanoparticles and thin films.

This work also explores the experimental parameters for obtaining good data for reliable ePDFs from nanomaterials.  Principally, samples should be thin enough or, for the case of nanoparticles, have a sufficiently small diameter.  What this diameter is depends on the average atomic number of the sample.  For Au, 100~nm diameter NPs gave significant coherent multiple scattering, 2.7~nm thick films did not.  For all materials we expect that 10~nm and smaller particles will scatter kinematically; and these are precisely in the size-range of nanoparticles that benefit the most from a PDF analysis~\cite{masad;prb07}. Obtaining a good powder average is also a very important part of powder diffraction regardless of the probing technique, XRD, ND or ED. This can be easily achieved by using a large sample volume in ND and spinning the sample in XRD. However, in ED, both of these methods become difficult due to the limitations of the configuration and careful sample preparation in this regard is very helpful. Again, for the particular application in nanoparticle structure characterization, the small size of the particles means that better powder averages can be obtained even from small sample volumes.  However, the quality of the powder average should be checked by visual inspection of the ED images from the CCD, which is readily done as is evident in the figures in this paper.  The powder average can be improved by increasing the beam-spot size on the sample and also by taking multiple images from different regions of the sample and averaging them. The maximum Q$_{\max}$ attainable is determined by the operational energy, camera length, dimensions of the detector and the diameter of the microscope, but in general should be maximized. Our electron microscope configuration equipped with a CCD camera limited Q$_{\max}$ to $\sim (17-18)$~{\AA}. The advantage of using a higher Q$_{\max}$ is the better real space resolution that results in the ePDF.  However, standard microscope configurations naturally give sufficiently high $Q_{\max}$ values for most applications.  Thus there seems to be no impediment to the use of ED from standard laboratory electron microscopes for quantitative nanoparticle structural characterization using the PDF.

\section {Acknowledgements}

We would like to acknowledge helpful discussions with Michael Thorpe.  We also thank Jon Hanson for allowing access to the X7B beamline at NSLS, which is supported by DOE-BES under contract No DE-AC02-98CH10886. Work in the Billinge group was supported by DOE-BES through account DE-AC02-98CH10886.  Work in the Kanatzidis group was supported by NSF through grant DMR-1104965.




\vfil
\pagebreak
%
%
%
%
%
%
%
%
\begin{table}[h]

\caption{Refined parameters for 2.7~nm thick nanoparticulate Au film, $\sim 100$~nm diameter nanoparticles (NP) from ePDFs and from a gold nanoparticle sample from xPDFs.  The structure model is the fcc bulk gold
structure, space-group Fm-3m. It was not possible to measure the nanoparticle size from the
ePDFs as we were not able to calibrate the intrinsic $Q$-space resolution of the ED measurement
allowing us to separate the  instrumental resolution and particle size effects in the ePDFs.}

\label{tab:Au_Results}
\begin{tabular}{lllllllll}
                                                        & ePDF (film)        & ePDF (NP)                 & xPDF                             \\
\hline
Q$_{\max}$  ({\AA}$^{-1}$)                  & 15.25                 &   15.25                          & 15.25                           \\
  Fit range  (\AA)                              & 1-20                  &   1-20                            & 1-20                             \\
  Cell parameter  (\AA)                     & 4.075(3)            &   4.076(2)                      & 4.058(1)                      \\
  U$_{iso}$   ({\AA}$^{2}$)                & 0.033(4)            &   0.006 (3)                     & 0.014(1)                      \\
  Diameter (\AA)                               & $\sim$27$^*$          &  $\sim$1000$^{**}$                 &24.51(9)     \\
 Q-damp    ({\AA}$^{-1}$)                 & 0.095(5)            &   0.095(5)                      & 0.047(2)                      \\
  Rw  (\%)                                         & 17                     &   24                               & 20                                \\
\hline
\multicolumn{9}{p{\textwidth}}{
$^*${film thickness measured during deposition} \par
$^{**}$NP diameter estimated directly from the TEM image
}
\end{tabular}
\end{table}
%
%
\begin{table}[h]
\caption{\label{tab:NaCl_results} Refined parameters for nanoparticulate NaCl from ePDFs and for a bulk powder
of NaCl from the xPDF.  The structure model is the fcc rock-salt
structure, space-group Fm-3m. It was not possible to measure the nanoparticle size from the
ePDFs as we were not able to calibrate the intrinsic $Q$-space resolution of the ED measurement
allowing us to separate the  instrumental resolution and particle size effects in the ePDFs. }
\begin{tabular}{llllllll}
                                                     & ePDF                                        & xPDF                         \\
\hline
Q$_{\max}$  ({\AA}$^{-1}$)                & 13.6                                        & 13.6                          \\
Fit range  ({\AA})                       & (0.2-30)                                    & (0.2-30)                     \\
  Cell parameter ({\AA})                 & 5.62(2)                                     & 5.63(1)                     \\
  $U_{iso}$ - Na ({\AA}$^{2}$)           & 0.007(5)                                    & 0.027(1)                     \\
  $U_{iso}$ - Cl ({\AA}$^{2}$)           & 0.004(4)                                    & 0.016(1)                     \\
 Q-damp          ({\AA}$^{-1}$)          & 0.095(5)                                    & 0.06(1)                      \\
  Rw  \%                                 & 33                                          & 6                               \\
\end{tabular}
\end{table}

\end{document}